\documentstyle{mn}
\def\aj{AJ}                   
\def\araa{ARA\&A}             
\def\apj{ApJ}                 
\def\apjl{ApJ}

\def\aap{A\&A}

\def\mnras{MNRAS}

\def\nat{Nature}

\title[Cooling flows and AGN mechanical power]{Cooling Flows as a Calorimeter 
of AGN Mechanical Power}

\author[Churazov, Sunyaev, Forman and
B\"ohringer]{E.~Churazov$^{1,2}$, R.~Sunyaev$^{1,2}$,
W.~Forman$^{3}$, H.~B\"ohringer$^{4}$\\
$^1$ Max-Planck-Institut f\"ur Astrophysik, Karl-Schwarzschild-Strasse 1, 85741
Garching, Germany\\
$^2$ Space Research Institute (IKI), Profsoyuznaya 84/32, Moscow 117810, 
Russia\\
$^3$ Harvard-Smithsonian Center for Astrophysics, 60 Garden St.,
Cambridge, MA 02138 \\
$^4$ MPI f\"{u}r Extraterrestrische Physik, P.O.Box 1603, 85740
Garching, Germany
}

\pagerange{\pageref{firstpage}--\pageref{lastpage}}
\pubyear{2001}

\begin{document}
\maketitle

\label{firstpage}
\begin{abstract}

The assumption that radiative cooling of gas in the centers of
galaxy clusters is approximately balanced by energy input from a
central supermassive black hole implies that the observed X--ray
luminosity of the cooling flow region sets a lower limit on AGN
mechanical power. The conversion efficiency of AGN mechanical
power into gas heating is uncertain, but we argue that it
can be high even in the absence of strong shocks. These arguments
inevitably lead to the conclusion that the time averaged
mechanical power of AGNs in cooling flows is much higher than
the presently observed bolometric luminosity of these objects. 

The energy balance between cooling losses and AGN mechanical power
requires some feedback mechanism. We consider a toy model in which the
accretion rate onto a black hole is set by the classic Bondi
formula. Application of this model to the best studied case of M87
suggests that accretion proceeds at approximately the Bondi rate down
to a few gravitational radii with most of the power (at the level of a
few percent of the rest mass) carried away by an outflow.

\end{abstract}


%

\sloppypar

\section{Introduction}
The radiative cooling time of gas in the central parts of rich galaxy
clusters is much shorter than the Hubble time. Without an external
energy source, the gas must cool below X--ray temperatures forming
so-called ``cooling flows'' (see Fabian 1994 for a review). The fate of the
gas, after it has cooled, is not clear and so far no unambiguous evidence for a
massive repository of cold gas has been found. An excess photoelectric
absorption in the cooling flow clusters derived from ROSAT and ASCA data
(e.g. Allen 2000, Allen et al. 2001) can be interpreted as due to
accumulation of the cold gas by the cooling flows over their lifetimes. This
interpretation however has been questioned by B\"ohringer et al. (2002)
based on the XMM-Newton data.  
Recent XMM-Newton and Chandra observations suggest that either gas does not
cool below 
$\sim1-3$ keV or it is somehow hidden from us (e.g. Fabian et
al. 2001, B\"ohringer et al. 2001). One possibility, discussed for
two decades, is that energy produced by AGN activity in the massive
galaxy at the center of a cooling flow balances radiative losses of
the X--ray emitting gas (see e.g. Binney \& Tabor 1995). Such a
picture would require fine tuning of the AGN power used for heating
the gas since both too much or too little heating would imply either
progressive heating or cooling of the X--ray emitting gas. If,
however, such fine tuning is present, then it should be possible to
use the X-ray luminosity of the cooling flow to estimate the 
the central AGN power. The gas in cooling flows is optically thin (at
least in the continuum) and the heating efficiency of the gas by
AGN radiation (see Ciotti \& Ostriker 2001) is rather low. A more
efficient source of heat is input from mechanical power, which we
consider below.

Since the balance between radiative losses and gas heating by AGN
mechanical input requires some feedback mechanism, we consider a
simple model with the accretion rate onto a black hole set by the
classic Bondi formula. Bondi accretion (Bondi 1952) is a 
hydrodynamical solution of spherically symmetric gas accretion without
angular momentum onto a point mass. A supermassive black hole in an
elliptical galaxy containing hot X--ray emitting gas is one site
where accretion may actually proceed at the Bondi rate (Fabian \&
Canizares 1988). The gas density and temperature in these sources can
be derived from X--ray observations. Then, if the mass of the black
hole is known, the mass accretion rate can be estimated. The observed
luminosities of the AGNs in these galaxies however are typically
several orders of magnitude short of expected values, calculated using
a standard $\sim$10\% conversion efficiency from the rest mass of
accreted material to radiation (Fabian \& Canizares 1988). It was
therefore suggested (e.g. Fabian \& Rees 1995, Reynolds et al. 1996)
that a hot ion torus forms around a black hole and accretion proceeds
as an Advection Dominated Accretion Flow (ADAF; e.g. Narayan \& Yi
1994). If the accretion rate is much smaller than the critical
Eddington value then the ADAF solution predicts that a very small
fraction of energy is radiated away. Recent
theoretical developments of radiatively inefficient flows show that
outflows can arise (Blandford \& Begelman 1999) and/or an outward
convective transport of energy may occur (e.g. Narayan, Igumenshchev
\& Abramowicz 1999; Quataert \& Gruzinov 2000). Observations of supermassive
black hole in elliptical galaxies in frame of the ADAF model and role of
outflow are discussed e.g. by Di Matteo et al. (1999, 2000) and Loewenstein
et al. (2001).   For many sources, well-collimated jets are observed and
(model dependent) estimates of the total mechanical power of the jets are
much larger than  directly observed luminosities of the AGN and the jet itself
(e.g. Pedlar et al.  1990, Owen, Eilek \& Kassim 2000).

\section{Cooling flows heated by  mechanical AGN power}
According to radio observations (e.g. Burns 1990), a large fraction
($\sim $71\%) of cD galaxies at the centers of cooling flows shows
evidence for radio activity. This fraction is much higher than in
non-cooling flow galaxies and suggests a possible link between the
presence of cooling gas and the activity of a central supermassive
black hole. Given that $\sim 10^8$--$10^{10} M_\odot$ black holes are
likely to reside in these elliptical galaxies, one can imagine that
these supermassive black holes, accreting matter at a rate well
below the critical Eddington value, could provide the necessary amount of
energy to offset cooling. Models involving gas heating
by an AGN have been extensively discussed in the literature (e.g. Tucker
\& Rosner 1983, Binney \& Tabor 1995, Soker et al. 2001, David et
al. 2001, Jones et al. 2001). In the discussion below, we present
a qualitative picture based particularly on the ROSAT, Chandra and
XMM-Newton data on M87 and NGC 1275 (B\"ohringer et al. 1993, 1995, 2001,
2002, Fabian et al. 2000, Churazov et al., 2000, 2001).

We assume that most AGN power goes into an outflow and bubbles of hot,
high entropy gas are created near the central engine. We further
assume that direct microscopic heating (e.g. heating of the cluster
gas by the high entropy gas in bubbles via thermal conduction) is
inefficient. We are then left with the ``mechanical'' energy exchange
between the bubbles and the cluster gas. The most important question
here is what fraction of energy goes into sound waves which can escape
the cooling flow region thus reducing the heating efficiency. From
simulations of rising bubbles (e.g. Churazov et al. 2001, Br\"uggen et
al. 2002, see also Rizza et al. 2000, Quilis, Bower and Balogh 2001), one
can note that the characteristic dimensionless numbers 
like the Mach number $M=v/c_s$ or Froude number $F=v/\sqrt{gL}$ (where
$v$ is the characteristic velocity, $c_s$ is the sound velocity, $g$
is the gravitational acceleration and $L$ is the characteristic
length) are not very far from unity. This precludes an immediate
conclusion on the character of generated disturbances and on the fate
of the deposited energy. On the other hand the same argument suggests
that while generation of sound waves certainly occurs during all
stages of the lobe evolution, it is unlikely that the great bulk of
energy is channelled into sound waves and that these waves carry away
most of the energy (unless very special conditions are created).

\subsection{The Late, Passive Stage of Bubble Evolution}

Let us first consider the latest stages of bubble evolution. 
Buoyancy drives the bubbles upward with a subsonic velocity and the
bubbles maintain approximate pressure equilibrium with the ambient gas.
Initial energy deposition
$E_0$ required to (slowly) inflate the bubble is the sum
of internal energy of the bubble $1/(\gamma-1)PV$ and $PdV$ work done by the
expanding bubble on the cluster gas, where $P$ is the pressure of the
cluster gas, $V$ is the volume of the bubble and  $\gamma$ is
the adiabatic index of the hot gas inside the bubble. Thus 
$E_0=\gamma/(\gamma-1)PV$. The buoyancy force acting on the bubble is
$F=V\rho \frac{d\phi}{dr}$, where $\rho$ is the density of the cluster gas and
$\phi$ is the gravitational potential. Assuming that the  bubble mass is very
small the buoyancy force must be equal to the drag exerted on the rising
bubble. The integral of the drag force over distance characterizes the
mechanical energy exchange between the bubble and the ambient gas. Assuming
that the bubble expands adiabatically during slow motion through 
the cluster atmosphere in hydrostatic equilibrium one can explicitly calculate
how much energy was extracted from the bubble:
\begin{eqnarray} 
\Delta E=-\int V\rho \frac{d\phi}{dr} dr=E_0-\frac{\gamma}{\gamma-1}PV=E_0 [1-\left
(\frac{P}{P_0} \right )^{1-1/\gamma}], 
\label{entalpy}
\end{eqnarray}
where $P_0$ and $P$ are the initial and final pressure in the bubble
respectively. Thus during their upward motion, the bubbles deposit their
energy when crossing roughly a few scale heights.  As an example,
for $\gamma=4/3$ and a pressure profile derived from the XMM-Newton
data (B\"ohringer et al. 2001, Matsushita et al. 2002) half of the
energy is lost by a bubble after reaching a distance of $\sim20$ kpc
from the its origin. 

An approximate estimate of how much energy goes into sound waves
during the initial phase of the bubble's rise can be obtained assuming
that the bubble instantly accelerates to a terminal velocity $v$ and
then maintains this velocity. For a solid sphere, Landau \& Lifshitz
(1986) give the energy emitted in sound waves as $E\sim
\frac{\pi}{3}\rho R^3 v^2$, where $\rho$ is the ambient gas density
and $R$ is the radius of the sphere. This energy is less than a
quarter of the internal energy of the bubble $E_i\sim
\frac{1}{(\gamma-1)\gamma}\frac{4}{3} \pi \rho R^3 c_s^2$, where $c_s$
is the sound speed of the gas outside the bubble.  After the initial
acceleration, the motion of the bubble is subsonic and steady (as seen
in the simulations) and we do not expect that the drag is
predominantly due to the generation of sound waves and weak shocks. Actual
values of the drag estimated from simulations (Churazov et al. 2001)
also support this expectation. A large fraction of energy goes instead
into the kinetic and potential energy of the entrained gas, turbulence
in the wake of the bubbles, and possibly into internal gravity waves.

Given the steep entropy profiles in typical cooling flow clusters (the
temperature is rising with radius while the density is deacreasing), 
large masses of entrained gas (even if accelerated to a fraction of
the sound velocity) cannot travel a long distance with the rising
bubbles (see Fig.10 in Churazov et al. 2001) and therefore kinetic energy can
not efficiently ``leak'' outside the cooling flow region with the
entrained gas.

Internal gravity waves are trapped within the inner region because the buoyancy
frequency is a decreasing function of radius in clusters (e.g. Balbus
\& Soker 1990, Lufkin, Balbus, \& Hawley 1995). The buoyancy
(Brunt-V\"ais\"al\"a) frequency $N$ can be written as
\begin{eqnarray} 
N^2=\frac{g}{\gamma}\frac{{\rm d ~ln} s }{{\rm ~d} ~r~~~},
\label{bv}
\end{eqnarray}
where $s=P/\rho^\gamma$ is the gas entropy, $r$ is the radius, $g$ is
the gravitational acceleration, and $P$ is the pressure.  For M87,
the buoyancy frequency can be estimated directly using the gas
temperature and density profiles derived from the XMM-Newton data
(Bohringer et al. 2001, Matsushita et al. 2002) under the
assumption that the gas is in hydrostatic equilibrium. These
estimates give $N\sim$ $1.2~10^{-7}$, $3.5~10^{-8}$, $1~10^{-8}$ rad/y
at radii of 2, 10 and 50 kpc respectively. Thus, the frequency of
the internal waves excited in the inner region is larger than the
buoyancy frequency in the outer parts and the waves are reflected back
into the central region.

Finally, we note that for subsonic turbulence (created in the wake of
a rising bubble), the ratio of the rates of generation of sound waves
and dissipation is (Landau \& Lifshitz, 1986)
\begin{eqnarray} 
 \frac{\varepsilon_{s}}{\varepsilon_{d}}\sim \left (\frac{v}{c_s}
\right )^5 \ll 1,
\label{dis}
\end{eqnarray}
where $v$ is a characteristic velocity and $c_s$ is the sound
speed. 

Note also that, as pointed out by Pringle (1989), sound waves
themselves are focused into the central region due to the decrease in
the sound speed at small radii. This effect may enhance the
dissipation of the generated sound waves within the central cooling
region. In the presence of steep temperature gradients, sound waves
traveling at large angles to the radius may even be  trapped in the cooling
flow region. The condition for trapping requires the ratio of the
square of the radius to the temperature, $r^2/T(r)$, be a decreasing
function of radius. However the
observed change of temperature (and associated change of the sound
velocity) is rather modest for M87 (see e.g. B\"ohringer et al.
2001) and the condition for trapping is not satisfied. The effect of
focusing is also modest and this mechanism is unlikely to strongly
affect the energy balance. 
 
Thus after the gas passes  through a phase of complex rearrangements
and mixing,  a significant fraction of the initial bubble energy  is
dissipated within the cooling flow region. Although sound waves are
inevitably generated  during this process, they can carry away only a
fraction of the energy comparable to that which is dissipated.

\subsection{Active, Early Stage of Bubble Formation}
At earlier stages of a bubble's (radio lobes) formation, shocks are
likely to occur. If the energy injection occurs in a small region of
undisturbed cluster gas then strong shocks will be sent into the
medium.  Strong shocks are efficient in energy dissipation and
there would be direct heating of the gas as it passed through
any strong shock. In a spherically symmetric cluster atmosphere,
with a density profile typical of a cooling flow ($\rho \sim 1/r$),
the strength of the shock decreases and it evolves into a sound
(compression) wave which can carry away some part of the shock
energy. Strong shocks leave behind relatively compact regions of high
entropy gas which contain significant fractions of the energy. The
numerical value of this energy fraction depends on the gas adiabatic
index and the radial profile of the ambient gas parameters. However,
an analogy with the Sedov solution for a strong explosion suggests
that the fraction of the shock energy deposited is ``of the order of
several times 10\%'' (Zeldovich \& Raizer 1966). The further
transfer of energy from this localized high entropy region to the rest
of the cluster gas follows the same processes as already described above.

Weak shocks on the other hand do not dissipate energy efficiently and
they may be the major source of energy leakage from the cooling flow
region. When the lobe passes through the transonic
expansion stage, a significant fraction of the injected energy can escape the
cooling flow region. But even during this stage, the fraction of energy which
remains in the form of internal energy within the lobe (and potential
energy of the displaced gas) is unlikely to be much less than half
the total.

The bottom line is that the fraction of mechanical energy deposited by
an AGN and dissipated {\it within the cooling flow region} is
high. For simplicity in the discussion below we assume that it is
equal to unity.

\section{Bondi Accretion and the Cooling Flow Luminosity}
The assumption that the whole cooling flow region is in a state of
quasi-stationary equilibrium requires the injected mechanical power 
to match the radiative cooling losses of the thermal gas.
 Therefore a feedback mechanism is required which must provide global
stable equilibrium 
between heating and cooling. We discuss below a toy model in which the
classic Bondi formula regulates the accretion rate onto the black hole
and provides a link between the gas parameters and the mechanical
power of an AGN.

The rate of spherically symmetric adiabatic gas accretion  without
angular momentum onto a point mass can be written as (Bondi 1952):
\begin{eqnarray} 
 \dot{M}=4\pi \lambda (GM)^2 c_s^{-3}\rho ,
\label{bondi}
\end{eqnarray}
where { $\lambda$} is a numerical coefficient, which depends on the
gas adiabatic index $\gamma$ (for { $\gamma=5/3$} the maximal valid
value of $\lambda$ allowing steady spherically symmetric solution is
$\lambda_c=0.25$), $G$ is the gravitational constant, $M$ is the mass
of the black hole, $c_s=\sqrt{\gamma\frac{kT_e}{\mu m_p}}$ is the gas
sound speed and $\rho$ is the mass density of the gas. Adopting
$\lambda=\lambda_c=0.25$ and substituting a reasonable set of parameters
one can write:
\begin{eqnarray} 
 \dot{M}_B=10^{-2} \left (\frac{M}{3\times10^9 M_\odot} \right)^2 \left (
\frac{n_e}{0.1~cm^{-3}} \right ) \left ( \frac{T_e}{1~keV} \right
)^{-3/2} ~ M_\odot~y^{-1}
\label{bondim87}
\end{eqnarray}
Here we use parameters appropriate for M87: $M=3\times10^9 M_\odot$
(Ford et al. 1994, Macchetto et al. 1997), typical (minimal)
temperature near the center $T_e\sim 1~keV$ and characteristic central
density $n_e\sim 0.1~cm^{-3}$ (B\"ohringer et al. 2001, Matsushita et
al. 2002). Given the angular resolution of XMM-Newton these values
correspond to 
the region of 1-2 kpc in size. A quick check of the Chandra M87 images shows
that the X--ray surface brightness is not steeply increasing from $\sim$1.5
kpc down 
to 0.3-0.2 kpc, where the emission from the nucleus and the jet start to
dominate. This suggests that the XMM-Newton data provide a reasonable estimate
of the density and temperature in the core region. The cooling
time of the gas with these parameters ($t_{cool}\sim \frac{3/2 P}{n_e^2
\Lambda(T)}\sim 10^7$ -- $10^8$ y, where $P$ is pressure and $\Lambda(T)$ is
a cooling function) is much longer than the characteristic accretion flow
time scale ($t_{flow}\sim \frac{R_B}{c_s} \sim$ few $10^5$ y) near the Bondi
radius: 
\begin{eqnarray} 
 R_B=2GM/c_s^2=0.1 \left (\frac{M}{3\times10^9 M_\odot} \right) \left (
\frac{T_e}{1~keV} \right )^{-1}~ kpc 
\label{rbondi}
\end{eqnarray}
Therefore (see e.g. discussion by Nulsen and Fabian 2000) the flow can be
treated as nearly adiabatic well outside the Bondi radius thus validating
the usage of Bondi formula (4).  

The above estimate of the accretion rate is a factor of 150 lower
than a similar estimate made by Di Matteo et al. (2000). This is at
least partly attributable to a lower density and a higher (adiabatic)
sound velocity adopted here.

Assuming that it is Bondi accretion which is actually powering an AGN
one can obtain an approximate estimate of the accretion efficiency:
\begin{eqnarray} 
 \eta = \frac{L_X}{\dot{M}_B c^2} \approx 0.05 \left (
\frac{\dot{M}_B}{0.01~ M_\odot~y^{-1}} \right
)^{-1} \left ( \frac{L_X}{3\times10^{43}~erg/s } \right ) 
\label{eff}
\end{eqnarray}
Here we adopted the X--ray luminosity of the cooling flow region in
M87 of $3\times10^{43}$~ergs s$^{-1}$ (Peres et al. 1998). From the above
expression one can conclude that if (i) the mass accretion rate is
approximately characterized by the Bondi formula and (ii) a large fraction
of the energy is carried by an outflow which is eventually dissipated into
heat in the cluster gas then a few percent of the rest mass of accreted
material has to be channeled into a jet or outflow. If the density at the Bondi
radius is a factor of $\sim$10 higher as suggested by Di Matteo et
al. (2000), then the efficiency will be lower, possibly as small as $\le$
1\%. Such an efficiency is still very high and simple outflow (Blandford
\& Begelman 1999) or convection dominated flows (Narayan, Igumenshchev \&
Abramowicz 1999) where most of
the material does not reach the innermost region cannot easily account
for such a high  energy ejection efficiency (if the accretion rate at
the outer boundary is set by the same Bondi rate).

The Bondi accretion rate is proportional to the ratio
$\frac{n_e}{c_s^3}\propto \frac{n_e}{T_2^{3/2}}\propto s^{-3/2}$, where $s$
is the entropy of the gas at the Bondi accretion radius. Thus for a given
black hole mass and cooling flow luminosity,  the entropy of the
accreting gas should have a certain value so as to provide the same amount of
energy which is emitted in X--rays:
\begin{eqnarray} 
 s \approx 3.5 \left (\frac{M}{10^9  M_\odot} \right
)^{4/3} \left (\frac{\eta}{0.1} \right
)^{2/3}   \left ( \frac{L_X}{10^{43}~erg/s } \right )^{-2/3} ~ kev~cm^2
\label{entropy}
\end{eqnarray}
A lower/higher entropy than this value implies too much/little heating 
to offset gas cooling  over the whole cooling flow region. Interestingly
(but naturally given that this is just another form of equation (\ref{eff}) ),
this value of entropy is close to that found in the cores of cooling
flows. Thus energy input is controlled by the minimum value of the gas
entropy in the cooling flow region. 

It should be stressed that although observed amounts of cold gas and
star formation rates in the cD galaxies are lower than the values
expected for the mass deposition rate inferred from the X-ray data,
they are clearly enhanced in the central cooling flow galaxies. This
suggests that some fraction of the gas is able to cool down, well
below X-ray temperatures. The presence of cold gas in the central
region may invalidate our simple estimates of the Bondi accretion rate
based on the X-ray determined density and temperature of the gas.  

\section{Discussion}
We speculate below on the implications of the assumptions made in the
previous two sections:
\begin{itemize}
\item {\bf A} The time averaged mechanical power of an AGN is approximately
equal to the radiative losses of the whole cooling flow region. 
\item {\bf B} The accretion rate given by the Bondi formulae can be used as an
(outer) boundary condition near the Bondi radius.
\end{itemize}

\subsection{Efficiency of Accretion}
Comparison of the accretion rate at the Bondi radius
(eq. \ref{bondim87}) and the observed cooling flow luminosities
implies (at least for M87) a high accretion efficiency -- of the order
of a few percent of the rest mass (see eq. \ref{eff}). Therefore
accretion should proceed with a rate $\dot{M}\approx const \approx
\dot{M}_B$ from the Bondi capture radius $R_B$ down to a few
gravitational radii. An outflow therefore should start very close to
the black hole. This in turn can be considered as an indirect argument
in favour of an outflow in the form of a jet. Interestingly, estimates
of jet mechanical power in some well studied cases (like M87 or
NGC1275) give very high values (e.g. Pedlar et al. 1990, Owen, Eilek
\& Kassim 2000) comparable to the total X--ray luminosity of the
cooling flow region. Pushing this speculative argument to the limit,
we can conclude that since our assumptions {\bf A} and {\bf B} mean
that the mechanical power of the AGN should adjust to the radiative
losses, then the jet power depends on the accretion rate and therefore
accretion is the major source of a jet's power in these objects. An
alternative possibility is that accretion serves as a trigger for jet
formation and the time averaged power is regulated through the duty
cycle of the AGN.

Numerical simulations of cluster formation (e.g. Frenk et al.,
1999) suggest that successive mergers may lead to  gas bulk
motions with typical kinetic energies at the level of $\sim 15$\% of the
gas thermal energy even for relatively relaxed clusters. The presence of bulk
motions in the cooling flow region implies that gas has angular
momentum.  If the gas captured at the Bondi radius has small, but nonzero
angular 
momentum, then at smaller radii the character of the accretion flow
may change significantly from the spherically symmetric Bondi
solution. The behavior of the flow in this region would depend on the
effective viscosity and the angular momentum transport and the flow may settle
on the geometrically thin (cold) or geometrically thick (hot)
solutions. Although the mechanical power, required by our assumption
{\bf A} is large, the  observed luminosities (e.g. from M87 or NGC
1275) are much lower than the standard $0.1 \dot{M}c^2$. Therefore an
ADAF-like solution is needed in this picture (at least in the inner
region) and the standard geometrically thin disk solutions (Shakura \&
Sunyaev 1973) extending down to the marginally stable orbit can be definitely
excluded (Fabian \& Canizares 1988, Fabian \& Rees 1995). We note here
that this problem is not only a consequence  of our assumptions, but
directly follows from the the comparison of jet power estimated by
Pedlar et al. (1990) or Owen, Eilek \& Kassim (2000) and observed
luminosities of the central objects in M87 or NGC1275. 

\subsection{Global Stability}
In the absence of a heat source, the total luminosity of the cooling
flow region is defined by the conditions (density and temperature) at
the cooling radius, i.e., at the radius where the cooling time
approximately equals the lifetime of the cluster. If the mechanical
power of the AGN fed via Bondi accretion is indeed the source of
energy which balances the cooling of the gas, then the actual AGN
power depends on the entropy of the gas near the Bondi capture radius,
i.e. at a distance of $\sim 0.1$ kpc from the central black hole. On
dynamical time scales, the cooling flow region adjusts to have the
lowest entropy gas at the bottom of the potential well. Too low values
of the entropy would increase the mechanical power of the AGN above
the radiative losses of the whole region and eventually will increase
the minimum entropy present in the system up to an equilibrium
value. Too high values of the entropy would mean that the power of the
AGN is too low to offset global radiative losses and eventually low
entropy gas will appear in the central region and the power of the AGN
will increase.

\subsection{Quasi-continuous Injection}
In the model discussed above, the time averaged AGN power is fixed by
the cooling flow luminosity. However the appearance of the central
region of the cooling flow may depend on the amplitude of the AGN
power variations.  For example, for very powerful but rare flares,
strong shocks are driven into the cluster gas and large bubbles should
be formed with a distinct envelope of hot, compressed gas (e.g. Heinz,
Reynolds \& Begelman 1998). Recent Chandra and XMM-Newton data,
however, do not show widespread evidence for shocks in the cooling
flow region (e.g., Fabian et al. 2000, McNamara et al. 2000, B\"ohringer et
al. 2001, 2002, David et al. 2001). These results rather favour a
scenario with more quasi-continuous injection of energy. In this case
(at least for spherical geometry), strong shocks should be confined to
spatially smaller regions. There are, however, examples where the
observed properties are best explained by strong shocks driven by
a powerful outburst -- the most spectacular one is perhaps NGC 4636
(Jones et al. 2001) in which both the surface brightness
morphology and the temperature structure suggest a strong shock. With
further observations of a large sample of objects it should be
possible to decide on whether sporadic or quasi-continuous energy injection
dominates.

\subsection{Sketch of the Cooling Flow Region}
Based on the above discussion, it is possible to outline the basic
properties of a cooling flow heated by buoyant bubbles, which are
quasi-continuously generated by an AGN. Bubbles are created in the
innermost region and are assumed to be filled with a hot, radio
emitting plasma. The size of a typical bubble is set by injection
power, gravity, and parameters of the cluster gas in such a way that
the lifetime of a bubble due to buoyancy is comparable with the
expansion time of the bubble due to injection power. For M87 and
NGC1275 this corresponds to a size in the range of few -- 10 kpc. The
bubbles have much larger entropy than the ambient gas and they rise
through the whole cooling flow region. Perhaps the most important
process operating in the innermost region is not a net heating, but
entrainment which continuously removes the low entropy gas from the
central region (Churazov et al. 2001, Nulsen et al. 2001). If initial
bubbles are created along some preferred direction, then bubble induced gas
motion may resemble the motion of air in the room with a heater: the higher
entropy gas rises along one direction while  lower entropy gas accretes
along others. This means that the radial entropy profile of a cooling
flow does not necessarily develop a flat core. 

When crossing a few scale heights, the bubbles i) lose their energy
through drag forces and ii) decrease in radio brightness
significantly. At this stage, bubbles can be identified as X--ray
surface brightness ``holes'' that are not associated with strong radio
emission (Churazov et al. 2000, Fabian et al. 2000, McNamara et
al. 2001). The dissipation of energy extracted from the bubbles is
distributed spatially with no obvious ``heating'' features 
e.g. shock waves. As seen from the simulations of the bubbles in M87,
the velocity with which the bubbles rise is smaller than the sound
velocity of the cooling flow gas, but not by a large factor. Therefore
the bubbles pass through the cooling flow during  time interval only
slighly longer than a sound crossing time $t_s$. This is at least
an order of magnitude shorter than a typical gas cooling time $t_c$.
An assumption that radiative losses of the gas is compensated by
the (efficient) dissipation of the bubbles energy then implies that
the total volume of the bubbles created during one cooling time of the gas
is approximately equal to the cooling flow volume (or in other words
the total energy of the bubbles is equal to the total energy of the
cooling flow). Therefore a volume filling factor of the bubbles within
the cooling flow region is approximately $t_s/t_c \ll 1$.

Thermal instabilities can take place on a background of  extensive
convective motions. Overdense (cool) lumps of thermal gas, entrained by the
bubbles and rising from (or falling back to) the central region might 
produce filamentary structures, stretching radially from the galaxy.  One
possible example of such a filament may be found in A1795 (Cowie et
al. 1983, Fabian et al. 2001). 

\section{Conclusion}
We show that even in the absence of strong shocks a significant
fraction of the energy of radio lobes inflated by an AGN is dissipated
in the cooling flow region. Only a comparable (but not overwhelmingly
dominant) fraction of the mechanical power of an AGN may escape the
cooling flow region in the form of sound waves.

If the the radiative cooling of the gas in the centers of 
galaxy clusters is balanced by energy input from a central
supermassive black hole,  then the observed X--ray luminosity of
the cooling flow region sets a lower limit on the AGN mechanical
power. The required power is much larger than the observed luminosity
of these AGNs, but is consistent with (model dependent) estimates of
jets power in some objects. 

We describe a toy model with the AGN power determined by Bondi accretion
for spherically symmetric accretion. Comparison of the accretion rate
predicted by the Bondi formula with the luminosity of the cooling flow
region in M87 suggests a high transformation efficiency  of 
rest mass of accreted material into the mechanical power of an outflow at the
level of a few percent. 

\section{Acknowledgements} 
WRF acknowledges support from NASA contract NAS8-39073. We are grateful to
the referee for important comments and suggestions. EC acknowledges useful
discussions with Marcus Br\"uggen, Marat Gilfanov,  Sebastian Heinz,
Hans-Thomas Janka and Alexey Vikhlinin.

\label{lastpage}
\end{document}